# Human Gender Prediction Based on Deep Transfer Learning from Panoramic Dental Radiograph Images

Isa Ataş

Computer Technology Department, Diyarbakır Vocational School of Technical Sciences, Dicle University, Diyarbakır 21100, Turkey

Corresponding Author Email: isa_atas@dicle.edu.tr

**ABSTRACT**

Panoramic Dental Radiography (PDR) image processing is one of the most extensively used manual methods for gender determination in forensic medicine. With the assistance of the PDR images, a person's biological gender determination can be performed through analyzing skeletal structures expressing sexual dimorphism. Manual approaches require a wide range of mandibular parameter measurements in metric units. Besides being time-consuming, these methods also necessitate the employment of experienced professionals. In this context, deep learning models are widely utilized in the auto-analysis of radiological images nowadays, owing to their high processing speed, accuracy, and stability. In our study, a data set consisting of 24,000 dental panoramic images was prepared for binary classification, and the transfer learning method was used to accelerate the training and increase the performance of our proposed DenseNet121 deep learning model. With the transfer learning method, instead of starting the learning process from scratch, the existing patterns learned beforehand were used. Extensive comparisons were made using deep transfer learning (DTL) models VGG16, ResNet50, and EfficientNetB6 to assess the classification performance of the proposed model in PDR images. According to the findings of the comparative analysis, the proposed model outperformed the other approaches by achieving a success rate of 97.25% in gender classification.

## 1. INTRODUCTION

Nowadays, artificial intelligence is actively used in a variety of domains, particularly in agriculture [1], [2], [3], health [4], [5], [6], [7] industry [8], [9], [10], natural language processing and voice recognition [11], [12], [13], security [14], [15], [16], generative networks [17], [18], remote sensing and hyperspectral imaging [19], [20], [21] etc. However, the modern methods developing rapidly in artificial intelligence technologies have shown remarkable success in image analysis and become more effective in medical applications. This study is based on the ground of the convolutional neural network, which has provided successful results in various image analyses with modern methods. Our dataset consists of 24,000 PDR images acquired from the local patients.

In the events where a significant number of mortalities occur as a consequence of natural disasters or catastrophes, making identification on human residue is necessary through the participation of professionals from various occupational groups. Due to the exposure of human remains to extreme and destructive external forces and their biological decomposition, making such identifications based on the existing remains becomes a challenging process [22]. The 2014 INTERPOL Disaster Victim Identification Standards emphasizes that DNA analysis, friction ridge analysis, and forensic odontology methods are the primary, most reliable, and efficacious identification techniques [23]. The determination of biological gender is the initial stage in the identification process.

Gender identification from human skeletal remains has been identified as an important factor in forensic science and bio-archaeology [24]. When determining gender in the defined areas, it is necessary to use as many methods or features as possible instead of using a single morphological feature as a reference [25]. In the literature, studies have been carried out for sex prediction with all existing bones of the human skeleton. mandible [26], calcaneus [27], metatarsal bone and phalanx [28], femur [24], patella [29], occipital condyle [30], hand bones [31] and sternum [32] are used to predict gender [33].

Despite the adversities, the mandible, which is usually known as the strongest, largest, and most resistant bone that remains intact, plays a significant role in gender prediction in forensic odontology [34], [35]. Therefore, the mandible, a skeletal component, is the focus.

Rather than taking a single morphological character as a reference, it is necessary to use as many procedures or features as possible while determining the gender of the unknown skeletal residues [25]. The PDR images with the mandible may provide information about dental status, age range, and gender. It is also conceivable to perform identification of a dead body or living individual whose identity is unknown, with such limited data on hand. Many diverse manual techniques are employed for gender prediction from PDR images of teeth. For instance, while an adult's skull is a reference material to identify gender with an accuracy of 80%, this accuracy can rise to 90% when the mandible is taken into account, additionally

[36], [37]. Although manual approaches are prone to error, the application of the aforementioned techniques requires a certain amount of time and experienced specialists (forensic anthropologist, pathologist, etc.) who are familiar with such techniques [38], [39].

As a result, the PDR images, acquired with fully automatic image techniques, covering the entire mandible and contributing positively to biometric identification; and thereby providing a holistic approach, were used for gender classification in our study.

## 2. RELATED WORK

It is possible to identify gender and age from skeletal remains in forensic medicine, osteology, and physical anthropology; however, gender determination by age is considered the most challenging issue [40]. The mandible reflects the anatomical distinctions between male and female individuals and poses sexual differentiation based on morphological characters [41]. The process of manually separating morphological characters for gender classification is common. Due to the intricacy of PDR images, researchers mostly concentrate on different morphometric and non-metric parameters of the mandible [42], [43], [44]. As a result of morphometric studies, Loth et al. achieved over 90% classification success as grounding on the characteristic of the ramus flexure, which is absent in males but is present in female individuals and is also regarded as a part of the mandible [45], [46], [47], [48]. Lin et al. set the upper limit of mandibular flexure, the maximum ramus vertical height, and upper ramus vertical height as discrimination parameters and correctly classified 81.7% to 88.8% gender classification among 240 three-dimensional mandibular models [49]. Deana et al. utilized numerous characteristics in non-metric studies, including jaw shape, eversion of the gonial angle, jaw profile, contour of the base of the mandible, and shape of the ramus, and reported a gender classification rate with an accuracy ranging from 75% to 95.2% [49]. Nagaraj et al. studied mandibular ramus flexure on a digital orthopantomogram. Statistical analysis was performed using SPSS software on the data of 100 subjects, and 71% accuracy was obtained [50]. In the cited manual studies in the literature, there were intact mandibles used without any pathology, loss of mandibular molars or abnormal molars, and teeth. Denis et al. proposed an automated solution for gender estimation based on deep learning techniques using convolutional neural networks from PDR images instead of employing only specific metric and non-metric indicators. However, as the test dataset advanced from 400 to 2000, the accuracy rate reduced from 96.8% to 92.3% [32]. Ivan et al. used deep convolutional neural networks (DCNN), which have attested successfully in image analysis, and achieved 94.3% accuracy in the test dataset used in DCNN models [36]. Based on the convolutional neural network developed by a multi-feature fusion module, Wenchi et al. suggested a new automated technique for gender estimation from panoramic dental x-ray images. With the method proposed, they attained 94.6% ±0.58% accuracy on the test dataset they used [38]. Ortiz et al. propounded a new technique for gender estimation through anatomical points that appear on panoramic radiographs using machine learning techniques. The accuracy rate was 68% for women and 74% for men [51]. The novel methods reported in the literature for gender estimation are mainly based on deep learning approaches and do not require any manual feature adjustment.

In addition to these studies, preprocessing techniques to improve the quality of dental images by years are presented in Table 1, various studies in dentistry that consider deep learning-based techniques are presented in Table 2 and Specific benchmarks in the advancement of Dental X-ray imaging methods are presented in Table 3 [52].

## 3. MATERIAL AND METHODS

### 3.1 Dataset collection and preparation

For binary classification of the PDRs, a dataset of images and label pairs was constructed and structurally tested by training in four alternative deep learning network architectures (VGG [53] – ResNet [54] – EfficientNet [55] – DenseNet [56]). This study examined a dataset of 24,000 PDR images from patients aged between 18 and 77 who received dental treatment in Diyarbakir Oral and Dental Health Hospital Periodontology clinic between 2015 and 2020. The female and male patient ratio in the data set was 58% and 42%, respectively. The images were captured using the Planmeca Promax 2D digital panoramic x-ray machine (anodic voltage 50-84 KV, current 0.5-16 mA, Planmeca, Finland) available in the clinic. The acquired PDR pictures had considerable differences in terms of contrast, location, and resolution parameters. That variation was one of the factors complicating gender identification. Figure 1 and Figure 2 illustrates sample female and male PDR images taken from a variety of patients. The PDR images were pre-processed to reduce complexity and focus on the mandible area. The histogram equalization method was applied to interpret the mandible area and teeth.

Furthermore, the resolution of original PDR images was resized from 3180 × 1509 to 224 × 224 pixels for the deep learning model without interfering with the aspect ratio value. An Open CEZERI Library (OCL) was utilized for pre-processing [57].

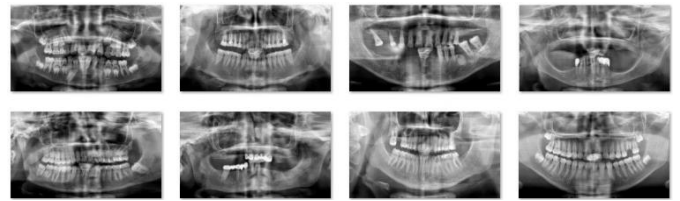

**Figure 1.** Sample PDR images of female patients by gender

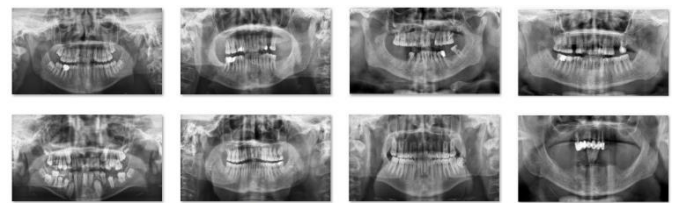

**Figure 2.** Sample PDR images of male patients by gender

Table 4 lists the distribution of the PDR image dataset 64%, 16%, 20% as training, validation and test set, respectively.

**Table 1.** Pre-processing techniques are addressed to recuperate the quality of dental images by year

| Author - Year | Relevant review findings | Detection/ Identification |
|---|---|---|
| Patanachai [58] - 2010 | The wavelet transform, thresholding segmentation, and adaptive thresholding segmentation are all compared. | Teeth detection |
| Frejlichowski [59] - 2011 | An automatic human identification system applies a horizontal integral projection to segment the individual tooth in this approach. | Human identification |
| Pushparaj [60] - 2013 | Horizontal integral projection with a B-spline curve is employed to separate maxilla and mandible | Teeth numbering |
| Lira [61] - 2014 | Supervised learning used for segmentation and feature extraction is carried out through computing moments and statistical characteristics. | Teeth detection |
| Abdi [62] - 2015 | Four stages used for segmentation: Gap valley extraction, canny edge with morphological operators, contour tracing, and template matching. | Mandible detection |
| Poonsri [63] - 2016 | Teeth identification, template matching using correlation, and area segmentation using K-means clustering are used. | Teeth detection |
| Zak [64] - 2017 | Individual arc teeth segmentation (IATS) with adaptive thresholding is applied to find the palatal bone. | Teeth detection |
| Mahdi [65] - 2018 | Quantum Particle Swarm Optimization (QPSO) is employed for multilevel thresholding. | Teeth detection |
| Fariza [66] - 2019 | For tooth segmentation, the Gaussian kernel-based conditional spatial fuzzy c-means (GK-csFCM) clustering algorithm is used. | Teeth detection |
| Aliaga [67] - 2020 | The region of interest is extracted from the entire X-ray image, and segmentation is performed using k-means clustering | Mandible detection |
| Rasool [68] - 2021 | Naive Bayesian (NB), Random Forest (RF) and Support Vector Machine (SVM) are used as classifiers for prediction. | Teeth detection |

**Table 2.** Presents various studies considering deep learning-based techniques in the field of dentistry

| Author - Year | Deep learning architectures | Detection/ Application | Metrics |
|---|---|---|---|
| Oktay [69] - 2017 | AlexNet | Teeth detection and classification | Accuracy |
| Chu [70] - 2018 | Deep octuplet Siamese network (OSN) | Osteoporosis analysis | Accuracy |
| Lee [71] - 2019 | Mask R-CNN model | Teeth segmentation for diagnosis and forensic identification | F1 score |
| Muramatsu [72] - 2020 | CNN (ResNet50) | Teeth detection and classification | Confusion Matrix |
| Rasool [68] - 2021 | Support Vector Machine (SVM) | Teeth detection and classification | Accuracy |

**Table 3.** Specific benchmarks in the advancement of Dental X-ray imaging methods

| Year | Dental Imaging Methods |
|---|---|
| 2005 | Level Set method |
| 2006 | Mathematical morphology & Connect component labelling |
| 2007 | Four field transformation & Support vector machine |
| 2008 | Clustering & Region growing |
| 2009 | Automatic iterative point correspondence algorithm & Hybrid knowledge acquisition |
| 2010 | Histogram based & Wavelet transform |
| 2011 | Canny, Sobel, Gaussian, Laplacian, Avarage filtering & Active contour |
| 2012 | Homomorphic filter, Distance, Adaptive windowing & Phase Congruency |
| 2013 | Harris operator, SVM Classifier & Gray-level co – occurrence matrix |
| 2014 | Bayesian Classifier, Gaussian filter & Local singularity analysis |
| 2015 | Cluster based Segmentation & Active shape model |
| 2016 | Fuzzy C-means & U-net architecture |
| 2017 | Neutrosophic Orthogonal Matrices, Transfer learning & Machine learning |
| 2018 | Multi-layer perceptron, Backpropagation algorithm & Deep learning based CNN |
| 2019 | Multilayer perceptron, Auto Regression model & Geodesic active contour |
| 2020 | Deep Convolution Neural Network |
| 2021 | Deep Convolution Neural Network & Transformer |

It is clear from the histogram graph that the number of patients aged between 25 and 50 are higher than the others.

**Table 4.** Training, validation and testing rates by gender in the PDR image dataset

| Class Label | Train (64%) | Validation (16%) | Test (20%) | Total (100%) |
|---|---|---|---|---|
| Female | 8,960 | 2,240 | 2,800 | 14,000 |
| Male | 6,400 | 1,600 | 2,000 | 10,000 |
| Total | 15,360 | 3,840 | 4,800 | 24,000 |

### 3.2 Transfer learning

Transfer learning is a deep learning method in which model parameters are used on a large pre-trained dataset. In other words, transfer learning is a method of machine learning where we reuse a previously trained model as a starting point for a model in a new task. Transfer learning is used in problems where there is not enough data for training or we want better results in a short time. Figure 3 shows the transfer learning procedure.

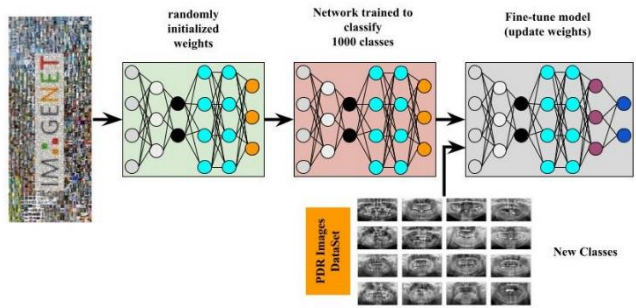

**Figure 3.** Transfer learning procedure

### 3.3 Proposed model

Convolution in the training of the convolutional neural networks and increment in subsampling steps causes a decrease in feature maps. However, gradient loss occurs in the image feature during transitions between cross-layers. The DenseNet architecture, in particular, was created to take advantage of the feature maps more effectively [56]. Each dense block in the DenseNet architecture has two convolution layers (conv), which are comprised of a varying number of repetitions. These are the 1x1 dimensional core defined as the bottleneck layer and the 3 × 3 dimensional core that will perform the convolution process. A 1x1 convolution layer is introduced before each 3 × 3 convolution layer to improve computational efficiency; thus, the number of input feature maps is reduced. However, each transitional layer contains a 1x1 convolution layer and a 2 × 2 average pooling layer with two strides [56].

In the classical CNN architecture, while each layer only has information about the feature map received from the previous layer, in the DenseNet architecture, however, each layer is updated with the inputs of all back layers. Since each layer is coupled feed-forwardly to others, any layer can access the feature information of all preceding layers. Reutilization of the feature map in dense blocks by different layers boosts the input and performance of the next layer, allowing for the generation of easy-to-train models. Figure 4 shows the comparison of the classical CNN model with the DenseNet model. When analyzing layer three in Figure 4, the DenseNet model is comprehended to receive information from all back layers, whereas the CNN model only gets input from the preceding layer, which is layer two. Using such a strategy improves the flow of information and feature maps in the DenseNet, resulting in a minimum loss. Several variants have been designed that belongs to the DenseNet family including DenseNet121, DenseNet169, and DenseNet201 [56].

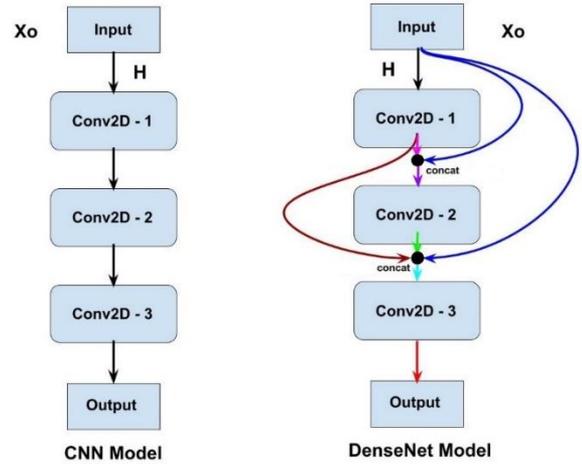

**Figure 4.** General CNN (left) and DenseNet (right) models

When the feature coupling is mathematized; if $X_0$ is defined as the input image, then the $H$ can be defined as a composite function consisting of three consecutive steps. In other words, $H$, the transfer function, consists of a combination of batch normalization (BN), rectified linear unit (ReLU), and 3 × 3 convolution (Conv.).

For the general CNN, while $l$'th output is generated by the $l-1$ 'th input,

$$X_l = H_l(X_{l-1}) \qquad (1)$$

In the DenseNet architecture, each layer concatenates the feature maps of previous layers and uses them as input for itself. Thus, $l$' th output is used as $X_0, \ldots, X_{l-1}$ input via taking the feature maps of all preceding layers and defining them as their assembly.

$$X_l = H_l([X_0, X_1, X_2, \ldots, X_{l-1}]) \qquad (2)$$

In the DenseNet, the size of the feature map expands as it passes through each dense layer and compiles existing features (the $k$ parameter) from previous layers. The growing rate indicated by the parameter '$k$' defines how dense architecture produces the most advanced outcomes. Thanks to the concatenate node built between the layers, the DenseNet performs well enough, despite having fewer parameters than the classical CNN architecture. The DenseNet121 model we proposed achieved better performance with approximately 7M parameters when compared to other models we tested in binary classification analysis. If every $l$'th layer of $H$ generates $k$ unit of the feature map, then $l$'th layer can be defined as:

$$k_l = k_0 + k \times (l-1) \qquad (3)$$

Here, $k_0$ refers to the number of channels in the input layer [56].

This study proposed a deep transfer learning strategy of the pre-trained DenseNet121 model to conduct binary classification from the PDR images. The proposed model was trained specifically with our PDR image set. The architecture of the DenseNet121 model is depicted in Figure 5.

The adjusted hyper-parameters of our model consisted of the learning rate, batch size, dropout rate, number of epochs, and optimizer. Table 2 lists the hyper-parameters used to train the proposed deep transfer model. In the performance analysis, the best accuracy rate with the minimum loss was attained using the values provided in Table 5. The Adam optimizer algorithm was used to optimize several DCNN models containing medical images [73].

**Table 5.** Selected hyper-parameters to train the proposed deep transfer model

| Hyper-parameters | Options |
|---|---|
| Learning rate | 0.0001 |
| Batch size | 16 |
| Dropout rate | 0.5 |
| Epochs | 50 |
| Optimizer | Adam |

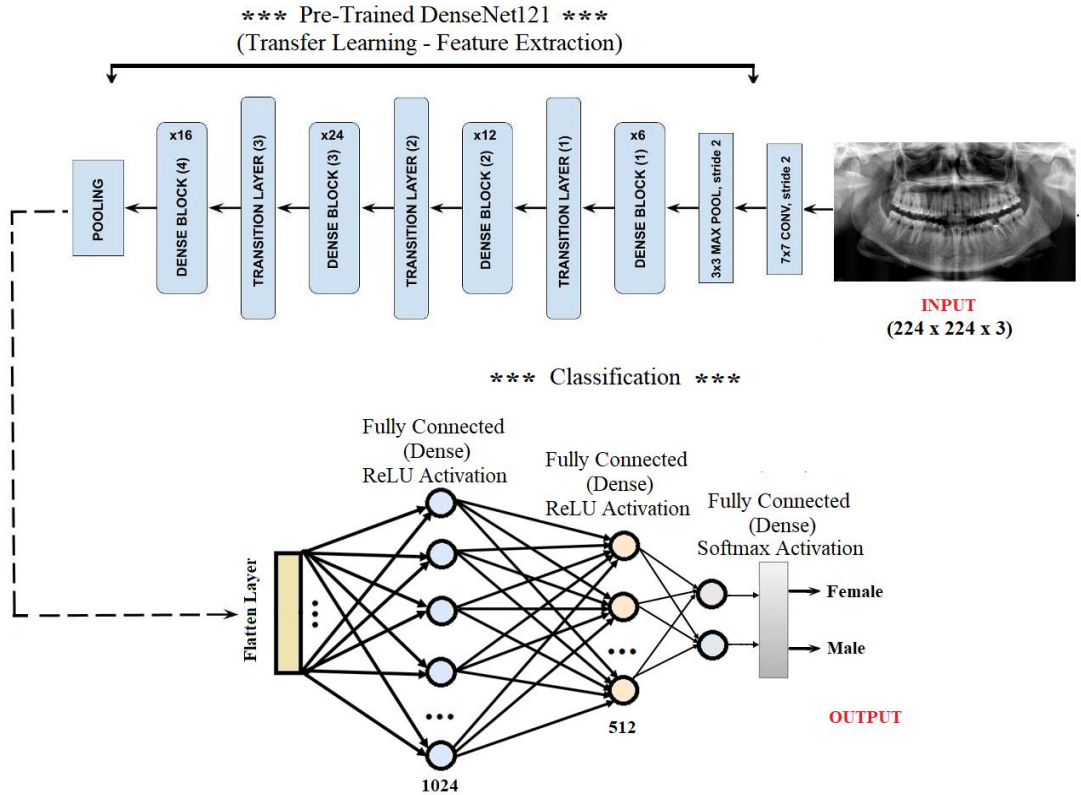

**Figure 5.** Architecture of proposed DenseNet121 deep transfer learning model for gender classification

In addition, when compared to other optimizers such as Stochastic Gradient Descent (SGD) [74] and RMSProp [75], the Adam optimizer had an appropriate function with minimal memory consumption and fast convergence. Our test dataset had approximately 4,800 PDRs, and such an amount could be regarded within the range of an adequate number to evaluate the performance of a gender estimation.

In our study, a 5-fold cross-validation technique was used to avoid overfitting or bias and to evaluate model training [76]. The 19,200 PDR images selected as the training dataset were randomly divided into five layers, and the dataset of each layer was sliced into 80 to 20 percent slices. The model was established using the train set in all steps and evaluated with the validation set. The statistical summary of the evaluation scores of the model was examined. An overview of the 5-fold cross validation performed in this study is presented in Figure 6. This process was repeated for each architecture (VGG16, ResNet50, EfficientNetB6 and DenseNet121) and the averages found as a result of the trials were reflected in the tables.

All models are trained and evaluated on Google Colab cloud platform [77] with 13,342 RAM - TeslaK80 GPU - NVIDIA T4 GPUs Card. The creation, training, validation and prediction of deep learning models are implemented using Keras [78] library and TensorFlow [79] backend engine.

### 3.4 Evaluation metrics

The confusion matrix is one of the most significant performance criteria in multi-classification problems. In this context, the accuracy, sensitivity, specificity, recall, and F1 score criteria are calculated through the confusion matrix [80]. The confusion matrix expresses the accuracy of the classifier by comparing the actual and predicted label values. Table 6 shows the general structure of a confusion matrix.

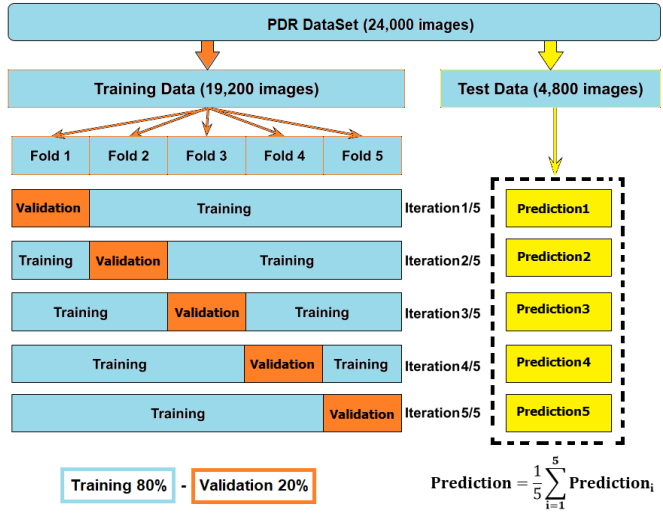

**Figure 6.** The overview of the performed 5-fold cross validation in this study.

**Table 6.** Confusion matrix

|  |  | **Predict Label** | |
|---|---|---|---|
|  |  | Female | Male |
| **Actual Label** | Female | True Positive (**TP**) | False Negative (**FN**) |
|  | Male | False Positive (**FP**) | True Negative (**TN**) |

True positive (TP) and true negative (TN) are where the model predicts the correct answer; false positive (FP) and false negative (FN) are where the model gets it wrong.

**TP:** Female data was estimated accurately and assigned as a true-positive label.
**FP:** Female data was estimated as Male and assigned as a false-positive label.
**FN:** Male data was estimated as Female and assigned as a false-negative label.
**TN:** Male data was estimated accurately and assigned as a true-negative label.

The accuracy refers to the ratio of correct (true) data defined to all data used. It is calculated as follows [80].

$$Accuracy = \frac{(TP + TN)}{(TP + TN + FP + FN)} \quad (4)$$

The precision refers to the ratio of positive data identified as true to all data identified as true. It is calculated as follows [80].

$$Precision = \frac{TP}{(TP + FP)} \quad (5)$$

The specificity refers to the ratio of negative data defined as true to the sum of negative data defined as true and positive data defined as false. It is calculated as follows [80].

$$Specifity = \frac{TN}{(TN + FP)} \quad (6)$$

The recall refers to the ratio of positive data identified as true to the sum of positive data identified as true and negative data identified as false. It is calculated as follows [81].

$$Recall = \frac{TP}{(TP + FN)} \quad (7)$$

F1 Score is calculated as the harmonic mean of sensitivity and recall, and it has a better measurement than the accuracy [80].

$$F1\ Score = 2 \times \frac{Precision \times Recall}{Precision + Recall} \quad (8)$$

## 4. EXPERIMENTAL RESULTS

In this section, the VGG16, ResNet50, and EfficientNetB6 models, commonly used as transfer learning models, were compared with the DenseNet121 model, which was recommended for binary classification from PDR images.

Table 7 illustrates the test accuracy values for the DenseNet models (121-169-201). With the highest accuracy value among the compared ones, the DenseNet121 model came to the forefront. Furthermore, the DenseNet121 was utilized as a reference model for the binary classification of PDR pictures due to its modest number of parameters. The accuracies of the proposed models were compared at various image resolutions for gender prediction, and the result was given in Table 8. The accuracy was significantly lower at the resolution of 96 × 96. Therefore, the 224 × 224 PDR resolution was preferred for comparative analysis in the study.

**Table 7.** Comparison of DenseNet models for test accuracy values

| Model | Total Parameters | Accuracy |
|---|---|---|
| DenseNet121 | 8,617,026 | **0.9725** |
| DenseNet169 | 14,880,322 | 0.9345 |
| DenseNet201 | 20,822,594 | 0.9467 |

**Table 8.** Comparison of the proposed model's accuracy at different image resolutions

| Model | Resolution of PDR | Accuracy |
|---|---|---|
| DenseNet121 | 96 × 96 | 0.8734 |
| DenseNet121 | 128 × 128 | 0.9233 |
| DenseNet121 | 224 × 224 | **0.9725** |

Table 9 shows the performance accuracy values for the test dataset of the four models compared. The selected hyper-parameters were employed in the training of four deep transfer learning. It was noteworthy that the accuracy value of the VGG16 was dramatically low. The DenseNet121 model, however, had the highest accuracy value among all the models.

The DenseNet121 architecture was compared with a different number of network layers. The results in Table 10 show that increasing or decreasing the number of network layers symmetrically has little effect on the accuracy of gender inference.

**Table 9.** Comparison of deep transfer learning models

| DTL Model | Input Shape | Total Parameters | Accuracy |
|---|---|---|---|
| VGG16 | (224, 224, 3) | 15,767,874 | 0.8220 |
| ResNet50 | (224, 224, 3) | 26,219,906 | 0.9260 |
| EfficientNetB6 | (224, 224, 3) | 43,855,505 | 0.9400 |
| DenseNet121 | (224, 224, 3) | 8,617,026 | **0.9725** |

The test inference times of the compared models is depicted in Figure 7.

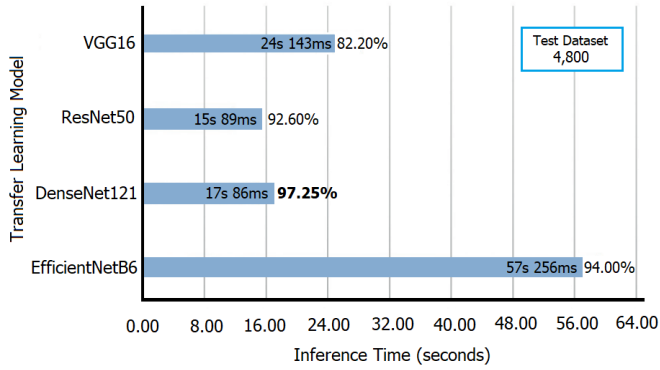

**Figure 7.** The elapsed inference times for the proposed transfer learning models

In the confusion matrix, the influence of false-positive and false-negative rates was presented in Figure 8. The Densenet121 model was found to generate minimal false-negative and false-positive results. For a visual representation of how successfully the DenseNet121 model identified samples for validation, the confusion matrix was employed. Table 11 shows the confusion matrix summary for the proposed model and previous deep learning-based PDR classification algorithms. The proposed model correctly classified 97.25% of the samples.

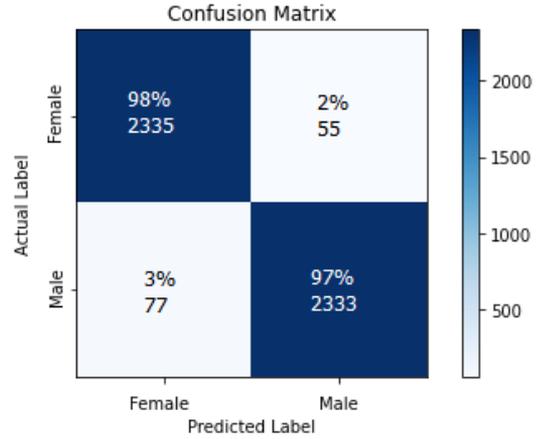

**Figure 8.** Confusion matrix analyses of the proposed model

When estimating the PDR image classification, the Gradient-weighted Class Activation Mapping (Grad-CAM) method reported in [82] was employed to determine sections to focus on for the identification process. The matrix generated by the filters in the last CNN layer of the proposed DenseNet121 model was superimposed on the actual PDR image for this purpose (Figure 9). The CNN feature map concentrating on the focus area is shown in the initial columns in Figure 9. The colors in the feature map refer to the convolutional neural network's targeted areas. The network's focus gradually increases as the color shifts from yellow to red. However, the secondary columns represent a coupling condition (high similarity score). The overlaid map in this column shows the focused areas with brighter yellowish and reddish colors spread over a wider area covering the maxillary, mandibular, and nasal sections. Considering the Grad-CAM and superimposed images examined, the proposed model is acknowledged to focus on the mandible and the teeth area and is a suitable and reliable instrument for forensic medicine practices.

**Table 10.** Backbone network configuration experiments

|  | DenseNet121-A | DenseNet121-B | DenseNet121-C | DenseNet121-D |
|---|---|---|---|---|
| feature_map_1 (Dense Layer) | 1024 | 1024 | 1024 | 1024 |
| dropout_1 (50%) | - | - | - | - |
| feature_map_2 (Dense Layer) | - | 512 | 512 | 512 |
| dropout_2 (50%) | - | - | - | - |
| feature_map_3 (Dense Layer) | - | - | 256 | 256 |
| dropout_3 (50%) | - | - | - | - |
| feature_map_4 (Dense Layer) | - | - | - | 128 |
| dropout_4 (50%) | - | - | - | - |
| accuracy: | 91.70% | **97.25%** | 95.45% | 93.00% |

**Table 11.** Testing analysis of the proposed and the other deep learning based PDR classification models

| **Model** | VGG16 | ResNet50 | EfficientNetB6 | DenseNet121 |
|---|---|---|---|---|
| **Precision** | 0.8075 | 0.8958 | 0.9355 | **0.9680** |
| **Recall** | 0.8262 | 0.9547 | 0.9447 | **0.9769** |
| **F1 Score** | 0.8219 | 0.9148 | 0.9390 | **0.9725** |
| **Specifity** | 0.8175 | 0.9074 | 0.9355 | **0.9680** |
| **Accuracy** | 0.8220 | 0.9260 | 0.9400 | **0.9725** |

A summary of the studies for gender estimation through PDR images is provided in Table 12. We could not compare previous studies and the study we proposed since there were neither the codes nor the datasets of the gender classification to access from the PDR images available in the literature.

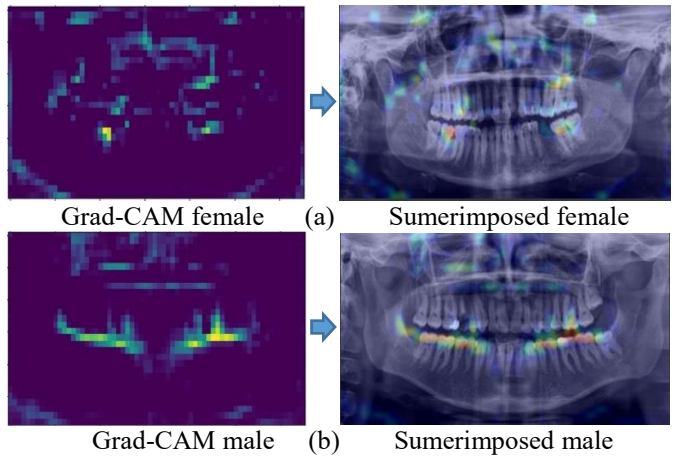

**Figure 9.** Visualization of the superimposed feature map onto PDR test image for both female and male cases

Figure 10 shows the performance graph of the training/test losses and accuracies of the DenseNet121 architecture for the 19,200 sampled training dataset. It was observed that the proposed model attained significant accuracy and loss values even in the 50th epoch.

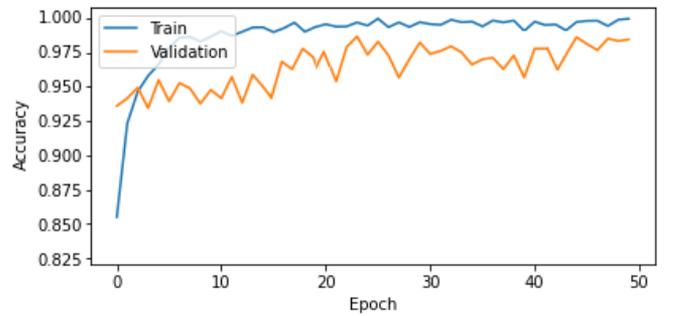

(a) Training and testing accuracy analysis

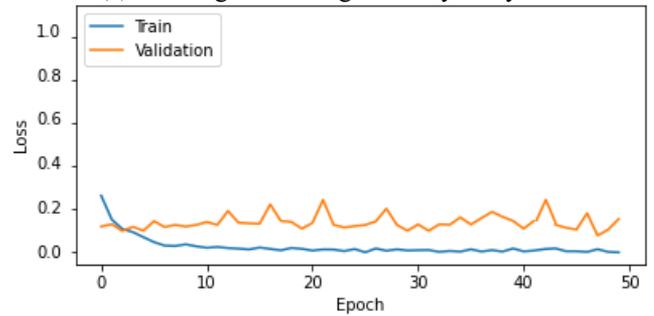

(b) Training and testing loss analysis

**Figure 10.** Training and validation analysis over 50 epochs

**Table 12.** Summary of studies for gender estimation from PDR images

| Author | Years | Total Dataset | Dental Imaging Methods | Accuracy |
|---|---|---|---|---|
| Steyn [84] | 2008 | 192 | Discriminant function | 79.7 - 95.4% |
| Jardin [85] | 2009 | 76 | Artificial Neural Networks, Metric methods | 68 - 88% |
| Saini [87] | 2011 | 116 | Mandible metric | 80.20% |
| Indira [88] | 2012 | 100 | Mandible metric | 76 % |
| Kim [86] | 2013 | 104 | Discriminant function | 65.4 – 89.4% |
| Nagaraj [50] | 2017 | 100 | Metric measurements | 71.00% |
| Deana [49] | 2017 | 128 | Metric measurements | 75.20 - 95.20% |
| Badran [25] | 2015 | 419 | Metric measurements | 70.90% |
| Oliveira [43] | 2016 | 160 | Metric measurements | 93.33 - 94.74% |
| Alias [89] | 2018 | 79 | Metric measurements | 78.50% |
| Denis [22] | 2019 | 4,000 | Convolutional Neural Network | 96.87% ± 0.96% |
| Ivan [36] | 2019 | 4,155 | Deep Convolutional Network | 94.30% |
| Wenchi [83] | 2020 | 19,776 | Multiple Feature Fusion | 94.60% ± 0.58% |
| Nicolás [90] | 2020 | 2,289 | Deep Neural Network | 85.40% |
| Mualla [95] | 2020 | 1,429 | Deep Neural Network | 95.80% |
| Rajee [91] | 2021 | 1,000 | Deep Convolutional Neural Network | 98.27% |
| Nithya [92] | 2021 | NAN | Deep Convolutional Neural Network | 95% |
| Rasool [68] | 2021 | 485 | Support Vector Machine | 92.31% |
| Santos [93] | 2022 | 1,142 | Library Support Vector Machine | 96% |
| Vila [94] | 2020 | 3,400 | Deep Convolutional Neural Network | 90% - 96% |
| **This study** | **2022** | **24,000** | **Deep Convolutional Neural Network** | **97.25%** |

## 5. CONCLUSIONS

Gender prediction is a critical and necessary process in forensic identification. Forensic experts and medical specialists employ traditional methods for gender estimation after years of training and education. In this study we proposed the DenseNet121 model using a deep transfer learning network and a fully automated technique to process panoramic dental x-ray images. The structural flexibility of the DenseNet121 architecture and the use of lesser parameters resulted in high-speed execution of training and verification processes. The weighted loss function was employed to eliminate the imbalance in gender classification, and the combination of early stopping and transfer learning was used to prevent over-learning. The best performance was achieved for the 4,800 test datasets with a classification accuracy of 97.25%. The proposed model, along with Grad-CAM based analysis also revealed that the mandible circumference and teeth are the most significant areas to consider in gender classification.